# Leveraging LLMs for Unstructured Claims Data Analysis

## Short Title: LLMs for Unstructured Claims Data

*by Robert D. Lieberthal, PhD;[1] Richard Tran, MD, MSCS;[2] Vietbao Phan, PharmD;[3] Jawand Singh;[1,4] and Elizabeth Sottung, PharmD[3]*

[1]Lieberthal & Associates, LLC; [2]MDSight, LLC; [3]Thomas Jefferson University; [4]William & Mary University

The Casualty Actuarial Society Artificial Intelligence Working Group provided funding and support for this research.

# Executive Summary

Actuaries traditionally rely on structured numerical and categorical data for reserving and ratemaking. Valuable predictive information in unstructured text formats such as medical records, adjuster notes, and call transcripts often remains unused. Manual processing of these documents is time-consuming, introduces reviewer-to-reviewer inconsistency, and faces significant scalability constraints. The purpose of this project is to develop a proof-of-concept approach to using large language models (LLMs) to extract structured predictors from unstructured claims data. The goal is to address the gap between rich narrative information and quantitative modeling requirements.

Medical records processing serves as the primary use case for this project. We implement a two-stage LLM processing architecture, separating document-level extraction (Stage 1) from claim-level synthesis (Stage 2). Each claim maintains its identifier, and documents are time-stamped to enable augmentation of existing datasets with LLM-derived insights. The modular four-script Python pipeline includes synthetic data generation and core processing, with standardized JavaScript Object Notation (JSON) schemas for integration with existing actuarial systems.

The framework uses Synthea Fast Healthcare Interoperability Resources (FHIR) bundles enhanced with LLM-generated synthetic documents to create complete claim ecosystems for testing. This ecosystem includes synthetic medical notes, phone calls, and adjuster assessments. Synthetic data allow for controlled testing with known ground truth characteristics and shareable, open-source datasets. The prototype extracts 36 actuarial variables that are used in reserving, ratemaking, and claims management functions.

We validate the results of the LLM extraction with a review by two independent clinician experts. We extract 14 core variables, including medical complexity, litigation risk, and claim development pattern. A scoring rubric based on a taxonomy that employs categorical values for qualitative assessments with Likert scoring (1–5) is the basis for expert review. Validation against synthetic data demonstrated overall extraction accuracy of at least 4 out of 5 across all variable categories, with strong confidence score calibration and a moderate degree of interrater reliability.

Finally, integration with chain ladder reserving shows the practical value of this approach. Severity-segmented analysis using LLM-extracted classifications shows the applicability of the approach to actuarial estimates of overall claims, and it produced ultimate estimates that lowered error from 6.5% to 4.0%, a 2.5 percentage point improvement. The integration also shows the interoperability of the LLM extraction with a wider set of actuarial methods.

The prototype includes working demonstration code, JSON output schemas, and a validated synthetic test dataset. The modular architecture supports phased deployment, enabling validation against real claims data while managing implementation risk. Future priorities include extension to additional insurance lines, integration with existing actuarial software platforms, and ongoing validation protocols to ensure that the deployment of LLMs results in accurate insights.

# 1 Introduction

## 1.1 Context and motivation

### 1.1.1 Background

Large language models (LLMs) have demonstrated remarkable capabilities across numerous business applications, from customer service automation to document analysis and decision support systems. The broader insurance sector has begun exploring these technologies for various operational functions, and the actuarial profession has been evaluating how these tools can enhance traditional analytical methods while maintaining the rigor and transparency that regulatory frameworks demand (National Association of Insurance Commissioners 2023; Balona 2024).

Actuaries primarily rely on structured numerical data for core functions such as reserving and ratemaking decisions, as well as for claims management. Unstructured claims data are incorporated into actuarial analysis after extensive manual processing by claims professionals and data management and governance teams. Processing includes the review of medical records, adjuster notes, and other narrative documentation to extract relevant information for coding into categorical variables. This human-intensive process is time-consuming, introduces reviewer-to-reviewer inconsistency, and faces significant scalability constraints. The volume of documents that can be manually reviewed is fundamentally limited by available human resources and associated costs, creating a gap between the rich narrative information generated during claims processing and the quantitative models that drive actuarial decisions.

LLMs represent a fundamentally different approach to processing unstructured text compared to earlier natural language processing (NLP) techniques. Rather than relying on rigid rule-based systems, LLMs employ neural network architectures trained on vast corpora of text to develop a sophisticated understanding of language structure, context, and meaning (Yang et al. 2025). For claims data processing, LLMs offer three valuable capabilities: natural language understanding that comprehends terminology and abbreviations without extensive manual coding, context awareness that distinguishes between different uses of the same term based on surrounding information, and the ability to handle varied document formats and writing styles through unified frameworks rather than requiring format-specific customization.

The potential applications extend across the full spectrum of actuarial functions. Systematically extracted variables describing injury severity and treatment complexity could enable more nuanced reserve estimates that reflect actual heterogeneity within claim portfolios. Features extracted from historical claims documentation could inform more accurate pricing, though such applications require careful evaluation. Early identification of high-complexity cases or litigation risk factors could support more efficient resource allocation and intervention strategies, leading to better claims management.

### 1.1.2 Literature review

The application of text mining and NLP to insurance claims data has evolved during the past two decades. Foundational work by Popowich (2005) established approaches for using NLP concept matching to extract semantic indicators from healthcare claims. Gharehchopogh and Khalifelu (2011) provided critical analysis distinguishing text mining from NLP approaches, highlighting that more than 80% of potentially useful business information exists in unstructured form. Zappa et al. (2021) demonstrated how text mining techniques can extract latent risk information from accident narratives to create new risk covariates for ratemaking.

Several developments in the insurance industry have both supported and raised cautionary notes regarding the use of artificial intelligence (AI) and LLMs. In 2023, the National Association of Insurance Commissioners issued a model bulletin titled *Use of Artificial Intelligence Systems by Insurers,* which established governance, documentation, and fairness considerations for systematic AI approaches (National Association of Insurance Commissioners 2023). Recent developments in LLMs have created new opportunities. Balona (2024) introduced ActuaryGPT, exploring LLM applications across various actuarial functions. Gilbert et al. (2024) demonstrated automated extraction from complex medical documentation, while Guo et al. (2024) established performance benchmarks for LLM-based health-related text classification. Richman (2024) provided direction for AI integration in actuarial practice, envisioning “AI-enhanced actuaries” who leverage these technologies while maintaining professional standards. However, Ressel et al. (2024) raised critical concerns about trust and the "uncanny valley" effect in AI applications for high-stakes insurance decisions.

## 1.2 Problem statement

The claims life cycle generates diverse unstructured data at each stage. This project focuses specifically on lines of insurance like workers' compensation with substantial medical claims. In these lines of insurance, medical records, adjuster notes, phone call transcripts, settlement documentation, claimant statements, and regulatory filings all are forms of unstructured data generated as part of the claims process. In general, almost any form of insurance will generate these or other types of unstructured or qualitative data. Each format presents specific challenges for the goal of transforming qualitative data, like text, into structured numerical, categorical, or binary data. Optical character recognition (OCR) errors in scanned documents introduce inconsistencies. Provider documentation standards vary substantially across healthcare systems. Claims professionals may use inconsistent, industry-standard, or company-specific terminology for similar claim characteristics. For example, a medical record describing "aggressive multimodal pain management including interventional procedures and intensive physical rehabilitation due to refractory symptoms" clearly indicates elevated treatment complexity and likely elevated cost trajectory, yet standardizing such descriptions into usable actuarial variables traditionally requires manual expert review.

The inability to systematically incorporate unstructured data leads to suboptimal outcomes across actuarial functions. Actuaries may lack the granular information needed for accurate claim stratification when relying solely on structured fields like diagnosis codes, missing crucial indicators of medical complexity and treatment trajectories. Valuable signals about causation types, safety compliance, and litigation indicators are present in adjuster notes but remain underutilized, resulting in less precise risk classification. Inefficient processing delays identification of claims requiring early intervention or specialized handling, creating bottlenecks that slow settlement processes and increase administrative costs. Timely prediction of emerging risks before they escalate is crucial in an industry in which litigation costs and the time value of money make faster decisions valuable.

## 1.3 Research gap, opportunity, and objectives

Despite a growing body of work on LLM applications and emerging regulatory frameworks, there exists limited peer-reviewed research providing systematic, reproducible methodologies for applying LLMs to actuarial variable extraction from unstructured claims data. Existing studies typically demonstrate proof-of-concept applications but lack systematic extraction frameworks for reliably transforming diverse documents into standardized actuarial variables, validation protocols for benchmarking accuracy and establishing confidence scoring, implementation guidance enabling actuaries to adopt LLM-based processing while meeting professional standards, and open-source implementations allowing the actuarial community to validate and extend methodologies. This gap is particularly significant given the regulatory emphasis on fairness, transparency, and explainability, requiring systematic approaches that generate audit trails, confidence measures, and documentation for regulatory compliance.

This research addresses the identified gap through three interconnected objectives. First, we establish a systematic, reproducible methodology for applying LLMs to actuarial variable extraction from unstructured claims data. Replicability includes documentation of design decisions, explicit protocols for each processing stage, and transparent handling of edge cases and limitations. Second, we translate technical LLM capabilities into practical tools that actuaries can integrate into existing analytical processes. Practitioner readiness extends beyond technical functionality to encompass clear documentation, intuitive interfaces, and integration pathways that align with actuarial workflows and professional standards. Third, we validate LLM-extracted variables and demonstrate that results can integrate seamlessly into traditional actuarial workflows. Integration requires demonstrating both technical compatibility and analytical value. Outputs must work with existing actuarial systems and methods as well as improve decision-making accuracy or efficiency compared to current practices.

One overarching objective is that the workflow design emphasizes practical usability through step-by-step guidance, enabling implementation without extensive technical expertise. The result is a working demonstration displaying fundamental capabilities of LLM-based claims processing, a simplified pipeline optimized for medical records processing workflows, and clear guidance for organizations interested in expanding these capabilities.

# 2 Methodology

This research project employs a modular, prototype-focused approach designed to demonstrate core LLM capabilities while establishing a foundation for future development and organizational adoption. The research design combines synthetic data generation with production-ready pipeline development, enabling thorough validation of key concepts while providing clear guidance for organizations interested in expanding these capabilities. The overall project approach is shown below in Figure *1*.

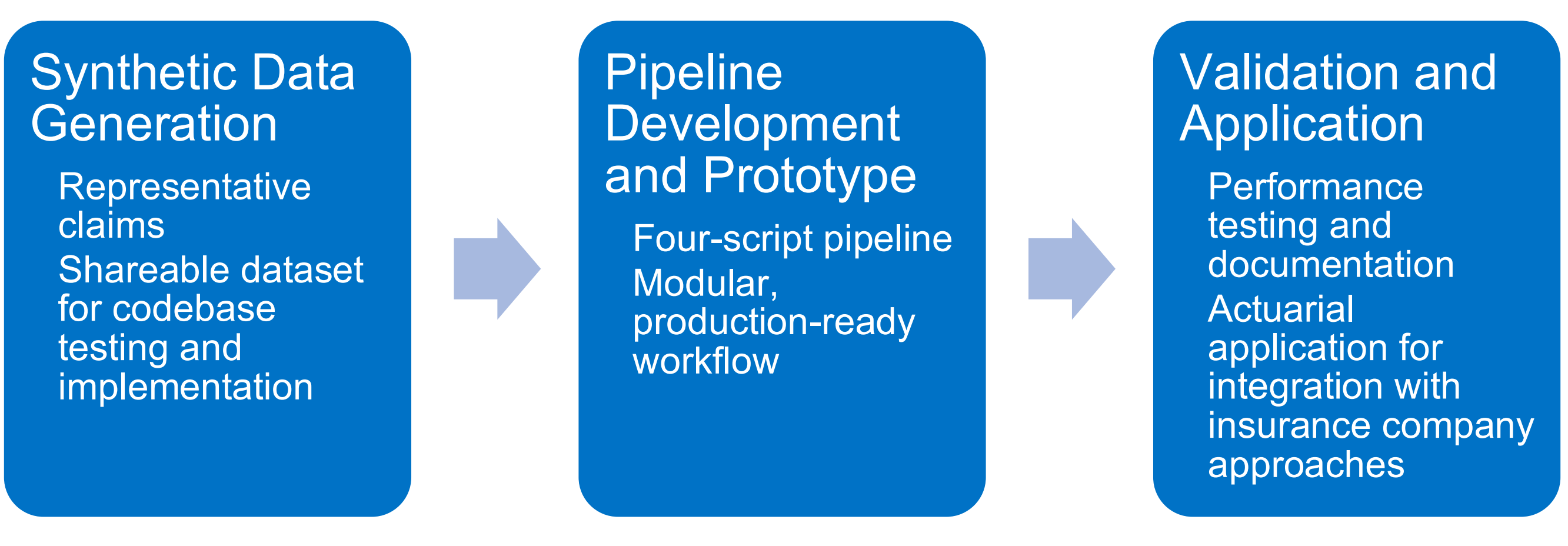

Figure 1. Project workflow and data processing pipeline.

## 2.1 Synthetic data approach

### 2.1.1 Data framework

The synthetic data generation leverages Synthea, an open-source synthetic patient generator that produces realistic Fast Healthcare Interoperability Resources (FHIR) medical records with known ground truth characteristics (Walonoski et al. 2018). Synthea generates complete patient medical histories including encounters, conditions, procedures, medications, and observations that follow realistic clinical progressions and temporal relationships. For this research, we configured Synthea to generate insurance scenarios, including injuries and exposures. An example of a synthetic worker's compensation case is shown in Example 1.

Example 1. Synthetic workers' compensation case.

> **Example Synthetic Case**
> Marcus Rodriguez, a 34-year-old concrete finisher with eight years of employment at Hillside Construction, sustained an L2 compression fracture on March 14, 2024. A gas-powered concrete vibrator malfunctioned during a commercial construction project, causing him to fall backward onto exposed rebar and then onto a concrete slab. Emergency department evaluation confirmed the L2 fracture with 25% anterior vertebral height loss, and the patient was fitted with a TLSO brace and referred to orthopedic spine specialist Dr. Robert Hayes for conservative management.

The Synthea output is processed through a two-script pipeline to generate a comprehensive claims ecosystem for testing. Script 1 (FHIR Bundle Processor) converts structured FHIR R4 medical data, a standardized format and application processing interface (API) specification for exchanging electronic healthcare information. It does so by using LLMs to transform key clinical elements (such as diagnoses, procedures, and medications) into narrative clinical notes, simulating a realistic medical documentation style. This LLM-generated narrative text becomes the input for subsequent analysis, as the original structured data are not preserved in the final output. Script 2 (Document Augmentation Engine) then enhances this medical data by generating necessary nonmedical synthetic documents, such as adjuster notes, phone transcripts, and settlement documents, creating a complete claim ecosystem.

Script 2 employs a two-stage master profile strategy. First, an LLM generates a comprehensive master case profile (~800+ characters) consolidating all claim details: patient demographics, incident specifics, injury descriptions, treatment timeline, and medical facts from FHIR encounters. This master profile then serves as shared context for generating each document type (adjuster notes, phone transcripts, medical letters, claimant statements, settlement notes) through document-specific LLM function schemas, ensuring factual consistency across all generated documents (e.g., identical claim numbers, incident dates, and medical details).

Each generated patient bundle contains clinical encounters spanning initial injury presentation through treatment completion or ongoing care, providing sufficient documentation complexity to test the extraction framework. The synthetic approach enables controlled testing where true variable values are known in advance, allowing precise measurement of extraction accuracy without requiring access to protected health information or expensive clinical expert review of real claims data during initial development phases.

### 2.1.2 Actuarial variable taxonomy and categories

The LLM extraction framework establishes a comprehensive taxonomy targeting 36 actuarial variables that may be useful in reserving, ratemaking, and claims management applications. This structure ensures LLM outputs align directly with established actuarial workflows, making them immediately integrable into existing analytical processes.

The core deliverable of the system focuses on extracting 14 key variables that provide immediate utility across three categories. While the complete taxonomy defines 36 variables to guide future development, the prototype implementation and validation concentrated on this 14-variable subset to demonstrate feasibility and establish extraction accuracy baselines. These include variables designed to support critical financial processes such as loss development analysis, reserve adequacy assessment, incurred but not reported (IBNR) estimation, risk classification, premium calculation, and experience rating applications. They also support operational efficiency by guiding resource allocation and settlement strategy.

The extraction framework employs multiple complementary strategies to mitigate LLM hallucination risks:

(1) Categorical constraints with controlled vocabularies limit qualitative factors (e.g., `injury_severity`, `litigation_risk`) to predefined values, including “unknown” when data are insufficient.
(2) Mandatory rationale fields require each extracted variable to cite specific document evidence and explain the reasoning, enabling human auditing.
(3) Multilevel confidence scoring provides document-level, variable-category, and overall confidence metrics (0–1 scale) to flag uncertain extractions.

(4) Actuarial consistency validation checks mathematical relationships (e.g., `ultimate_cost_prediction` must align with `ultimate_cost_category` range, `total_incurred_cost` cannot exceed predicted costs).

While these mechanisms constrain output form and detect internal inconsistencies, they cannot fully prevent coherent-but-incorrect extractions when the model misinterprets source documents.

The variable taxonomy captures qualitative factors typically inaccessible through structured data systems. Each of the 36 variables (detailed in **Error! Reference source not found.**) combines categorical assessment with validated numeric and Boolean values, with extracted results linked to claim identifiers and document time-stamps for integration with existing data warehouses and actuarial platforms. For operational deployment, additional production-grade controls would strengthen reliability: calibration analysis comparing confidence scores to empirical accuracy, systematic error categorization distinguishing missing information from misinterpretation versus true hallucination, and automated anomaly detection flagging statistically improbable extractions. The current implementation provides foundational quality controls suitable for research validation and proof-of-concept evaluation.

Table 1. Complete actuarial variable taxonomy (36 variables).

| Variable Name | Values |
|---|---|
| `Claim_severity*` | minor, moderate, major, catastrophic |
| `Injury_severity` | minor, moderate, major, catastrophic |
| `Medical_complexity` | simple, moderate, complex, highly_complex |
| `Treatment_type` | conservative, surgical, ongoing_chronic, complex_rehab |
| `Expected_development_pattern**` | fast_resolution, average, slow_development, long_tail |
| `Ultimate_cost_category` | <25K, 25K–50K, 50K–100K, 100K–150K, >150K |
| `Total_incurred_cost***` | Numeric (USD) ≥0 |
| `Estimated_outstanding_cost***` | Numeric (USD) ≥0 |
| `Ultimate_cost_prediction***` | Numeric (USD) ≥0 |
| `Permanent_disability_rating` | Numeric (%) 0–100 |
| `Recovery_timeline_days` | Numeric (days) ≥0 |
| `Maximum_medical_improvement_reached` | true, false |
| `Total_medical_costs` | Numeric (USD) ≥0 |

| Total_indemnity_paid**** | Numeric (USD) ≥0 |
|---|---|
| Settlement_amount | Numeric (USD) ≥0 |
| Reserve_adequacy***** | adequate, possibly_deficient, deficient, excessive |
| Medical_provider_quality | appropriate, questionable, excessive, inadequate |
| Risk_level | low, medium, high, extreme |
| Causation_type | equipment_failure, environmental_hazard, human_factor, multiple_causes |
| Industry_risk_category | construction_heavy, manufacturing_heavy, healthcare_hospital, etc. |
| Industry_risk_level | low_risk, moderate_risk, high_risk, hazardous |
| Safety_compliance | full, partial, none, unknown |
| Experience_modifier_impact****** | favorable, neutral, adverse, severe_adverse |
| Employer_size_category | small, medium, large, enterprise |
| Claim_frequency_indicator | first_claim, repeat_claimant, high_frequency_employer, unknown |
| Litigation_risk | low, medium, high |
| Settlement_likelihood | high, medium, low |
| Management_complexity | routine, moderate, complex, high_touch |
| Fraud_risk | low, medium, high |
| Time_to_closure_days | Numeric (days) ≥0 |
| Recommended_reserve | Numeric (USD) ≥0 |
| Work_relatedness_status | accepted, disputed, denied, under_investigation |
| Return_to_work_status | returned_full_duty, returned_light_duty, not_returned, permanent_disability, unknown |
| Claim_closure_status | open_active, open_inactive, closed_settled, closed_denied, pending_settlement |
| Subrogation_potential | none, low, medium, high |
| Claimant_cooperation_level | excellent, good, fair, poor, adversarial |

Note: The prototype validation focused on 14 core variables: `Claim_severity`, `Injury_severity`, `Medical_complexity`, `Treatment_type`, `Expected_development_pattern`, `Ultimate_cost_category`, `Total_incurred_cost`, `Ultimate_cost_prediction`, `Maximum_medical_improvement_reached`, `Risk_level`, `Causation_type`, `Litigation_risk`, `Settlement_likelihood`, and `Management_complexity`.

* Claim_severity uses weighted aggregation per defined feature importance (settlement: 0.98, medical: 0.95, early documents: lower weights). Settled claims reflect final assessment; open claims use most authoritative available.

** Definitions for this variable are configurable in the yaml-config file.

*** These three variables are mathematically related (ultimate = incurred + outstanding), but we retain all three because they (1) originate from different documents with varying confidence levels, (2) serve distinct actuarial functions (reserving, IBNR, development triangles), and (3) enable mathematical consistency validation to detect extraction errors.

**** We include approximately seven cost fields reflecting actuarial practice: breakdown by type (medical, indemnity, total), breakdown by status (paid, outstanding, ultimate), and settlement amounts for closed claims.

| These support different actuarial functions (reserving, settlement analysis, and loss development). Users can select relevant subsets based on use case. |
|---|
| ***** `Reserve_adequacy` extracts the adjuster's documented assessment from claim notes rather than having the LLM perform actuarial calculations. It aggregates explicit statements like "reserves appear adequate given current treatment plan" or "initial reserve may be insufficient based on surgical complexity." This variable feeds into the `recommended_reserve` calculation logic (e.g., flagging when reserves may need adjustment). |
| ******* `Experience_modifier_impact` evaluates the claim's likely effect using defined actuarial criteria (cost, severity, frequency) that typically drive modifier changes. The LLM applies our specified logic to assess impact, providing a claim-level risk indicator for early intervention. |

The 36 variables presented in Table 1 represent Stage 2 outputs generated through claim-level aggregation, not direct Stage 1 document-level extractions. The two-stage architecture operates as follows: Stage 1 (document extraction) processes each document independently (e.g., phone transcripts, adjuster notes, medical provider letters, settlement notes, claimant statements, clinical notes) to extract 62 document-specific features, such as `settlement_amount`, `total_incurred_cost`, `injury_type`, `treatment_protocol`, and `litigation_indicators`. Stage 2 (claim aggregation) then synthesizes these Stage 1 features across all documents for a given claim into the 36 final actuarial variables.

Variables such as `ultimate_cost_prediction` and `ultimate_cost_category` exemplify this synthesis: `ultimate_cost_prediction` aggregates document-level cost evidence (settlement amounts, incurred costs, medical expenses) with weighted importance based on document type and recency, while `ultimate_cost_category` derives categorical ranges directly from the numeric prediction to ensure mathematical consistency. Both Stage 1 document-level features and Stage 2 aggregated variables are accessible in the output; Stage 2 variables serve as the primary input for actuarial modeling, while Stage 1 features provide granular audit trails and document-specific context.

### 2.1.3 Selection criteria for medical records use case

Medical records processing serves as the primary proof of concept for this LLM framework, chosen strategically to demonstrate return-on-investment potential across multiple medical-driven claim types while addressing a critical gap in current actuarial practice. Successful processing of these records demonstrates immediate applicability to workers' compensation, medical malpractice, bodily injury, and other property and casualty lines where medical complexity significantly influences claim costs and outcomes. The selection of medical records as the primary use case reflects both their actuarial significance and their technical characteristics that make them suitable for demonstrating LLM capabilities.

The structured extraction of key medical information, including diagnosis codes, treatment complexity indicators, injury severity measures, and recovery timeline markers, enables actuaries to incorporate previously inaccessible clinical insights into their traditional modeling approaches. This focus on medical records addresses a critical gap where valuable predictive information contained in physician notes, treatment plans, and clinical assessments remains untapped due to the time and expertise required for manual review and coding. Current approaches to extracting information from medical records suffer from inconsistent categorization across different reviewers, incomplete capture of relevant information due to time constraints, and the inability to systematically process the volume of unstructured data generated throughout the claims life cycle. The approach directly addresses the actuarial profession's focus on evidence-based decision-making by providing systematic methods to incorporate clinical evidence and medical expertise into quantitative models.

## 2.2 System architecture (two-stage processing)

### 2.2.1 Architecture flow

The prototype implements a two-stage processing architecture designed to systematically extract actuarial variables from unstructured claims documents. This approach separates document-level extraction (Stage 1) from claim-level synthesis (Stage 2), addressing a fundamental challenge in claims processing: Individual documents within a claim file often contain overlapping, evolving, or contradictory information as shown in Figure 2.

The workflow design and integration strategy produces clear, reproducible processes from data ingestion to actuarial output while ensuring seamless integration with existing systems and methodologies. The workflow design emphasizes practical usability, providing step-by-step guidance that enables actuaries to implement LLM-based processing without requiring extensive technical expertise or significant modifications to established practices. The integration strategy addresses the critical need for compatibility through standardized JSON schemas, as seen in the center of Figure 2, cloud platform compatibility across major providers (e.g., Amazon Web Services (AWS), Microsoft, Google), and clear delineation of AI versus human roles in the analytical process, ensuring that LLM outputs can be readily incorporated into traditional actuarial tools and decision-making frameworks without disrupting existing workflows.

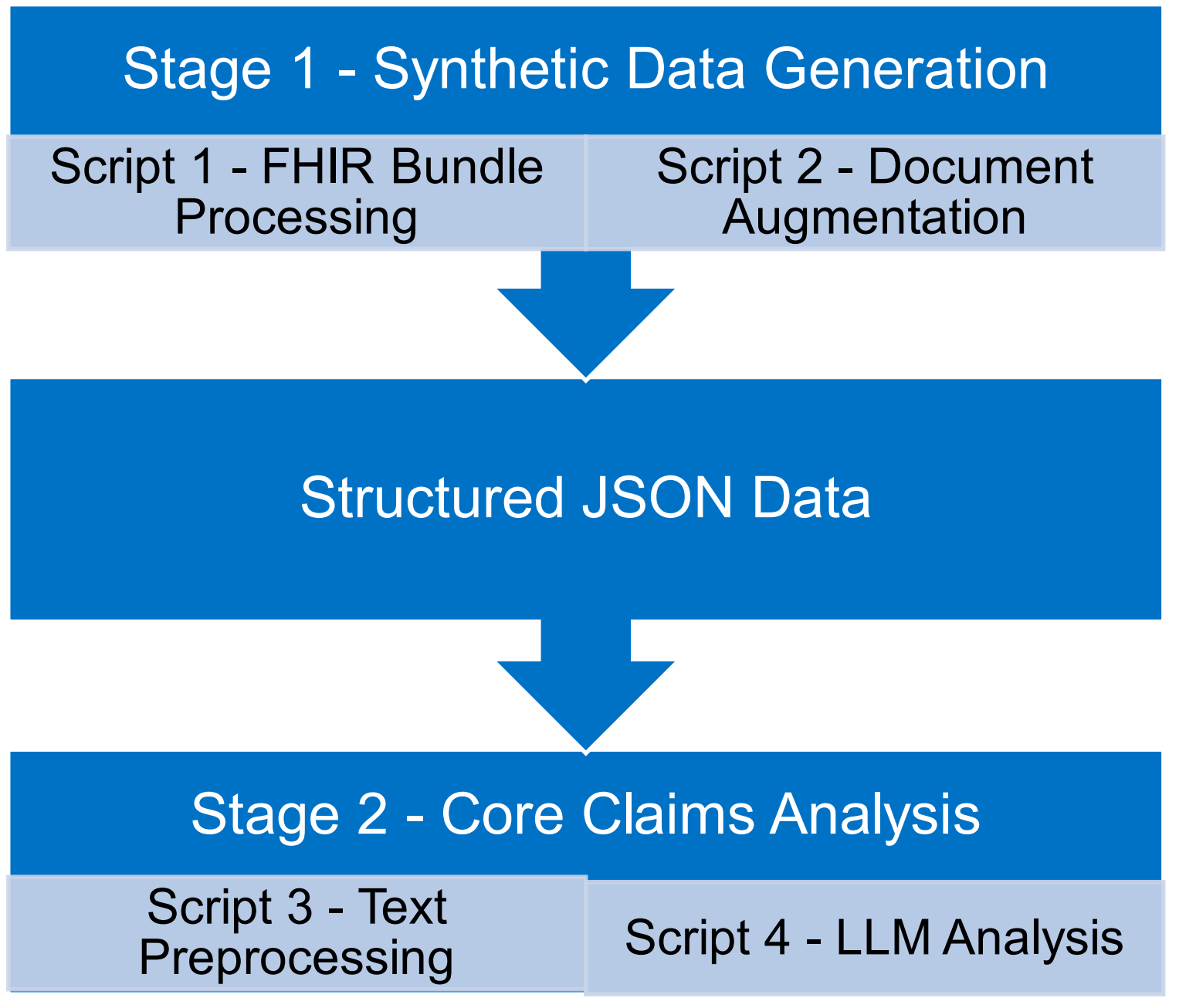

Figure 2. Integration workflow.

Stage 1 processes each document independently using document-specific prompts and extraction logic. Each document analysis produces feature extractions with confidence scores and supporting evidence. Stage 2 then aggregates all Stage 1 results for a claim, applies priority weighting by document type and recency, resolves conflicts through compound scoring, and generates final actuarial variables with comprehensive provenance documentation. The output is structured JSON containing claim-level variables organized by functional category (reserving, ratemaking, claims management) that is ready for integration into actuarial systems. Detailed architectural specifications, including the compound scoring formula and sequential processing implementation, are documented in the Technical Architecture and Methodology Document (TAMD) (see Appendix D).

### 2.2.2 Feature engineering framework

Feature engineering transforms narrative claims information into structured categorical variables suitable for actuarial modeling. The system converts unstructured narrative text into structured features through LLM-based natural language understanding. For example, a physician’s note stating, “Patient presents with severe lumbar radiculopathy, unable to ambulate without assistance” becomes the following:

- Injury severity: “major”
- Medical complexity: “complex”

These extracted features become variables that actuaries use to stratify claims, predict development patterns, and set appropriate reserves. The framework defines a comprehensive taxonomy of 36 actuarial variables (**Error! Reference source not found.**) organized into three functional categories: reserving, ratemaking, and claims management. The prototype implementation extracts 14 core variables selected for their immediate actuarial utility and extractability from typical claims documentation.

When multiple documents provide conflicting assessments of the same characteristic, the system resolves conflicts using priority weights that reflect document authority based on actuarial practice. Settlement notes receive the highest weight (1.0) as final assessments with complete claim knowledge, followed by medical provider letters (0.95) providing authoritative clinical opinions, initial adjuster notes (0.70) offering early professional assessments, and phone transcripts (0.45) capturing preliminary self-reported information. These weights are defined in configuration files based on established claims hierarchy rather than learned from data and were refined iteratively during prototype development.

Improper weights can significantly affect output quality. Categorical assessments such as severity classifications and litigation indicators show high sensitivity to weight errors because documents frequently contradict on subjective judgments. If early self-reports outweighed final medical assessments, the system would systematically misclassify claim characteristics. Cost predictions show moderate sensitivity because values typically converge as claims mature, while objective facts show minimal sensitivity as they rarely conflict across documents. The compound scoring mechanism (detailed in TAMD Section 3.3) combines priority weights with confidence scores and temporal decay to produce final values. Implementation details, including the script pipeline, prompt engineering approach, and confidence scoring mechanisms, are described in Section **Error! Reference source not found.**.

### 2.2.3 Data conversion

The technical implementation of data conversion requires attention to output specifications and integration requirements. The development of JSON schemas that align with existing actuarial data models and systems makes any solution standardized and more scalable. This is because standardization ensures that converted data can seamlessly feed into traditional actuarial tools and methodologies without significantly modifying established workflows. The conversion process also should include confidence scoring and uncertainty quantification, allowing actuaries to assess the reliability of automatically generated classifications and make informed decisions about when human review or additional validation may be necessary. Successful data conversion creates a bridge between the rich, unstructured information contained in claims documentation and the structured data requirements of actuarial analysis and modeling at a much lower cost in time and money than that of existing manual approaches.

## 2.3Document processing strategy

The system implements a comprehensive document processing strategy designed to handle the diverse formats, quality levels, and structural variations found in real-world claims documentation. This strategy bridges the gap between raw unstructured text and the structured extractions required for actuarial analysis. The prototype works with pre-deidentified datasets to focus on LLM methodology rather than privacy solutions, allowing the research to concentrate on core functionality. This constraint recognizes that actuaries, developers, and other users will be able to extend the prototype solution to a comprehensive solution for their individual companies. Ongoing model validation that regulators and other stakeholders may require as the claims language evolves can be added in the future as needed, including robust deidentification methodologies, Health Insurance Portability and Accountability Act (HIPAA)-compliant processing, and fine-tuned domain-specific LLMs trained on proprietary claims data.

### 2.3.1 Document categorization processing

The prototype processes six document types commonly found in workers' compensation and property-casualty claims:

- Clinical notes (physician visit documentation, examination findings)
- Claimant statements (first-person injury narratives, functional limitations)
- Medical provider letters (treatment updates, clinical assessments)
- Phone transcripts (initial claim intake conversations)
- Settlement adjuster notes (final evaluation and resolution documentation)
- Initial adjuster notes (investigation findings, liability assessment)

The architecture supports expansion to additional types including legal correspondence, independent medical examinations, and surveillance reports.

Input documents arrive as plain text files (.txt) with UTF-8 encoding, following a standardized naming convention that embeds metadata: `{patient_initials}_{claim_id}_{document_type}.txt`. The preprocessing pipeline normalizes text encoding, standardizes whitespace while preserving medical formatting, and extracts metadata from filenames and document headers.

Document types are identified through filename pattern matching using a document type mapping defined in the configuration. The system recognizes keywords in filenames and maps them to standardized document types. This explicit naming convention ensures reliable document classification without requiring content analysis.

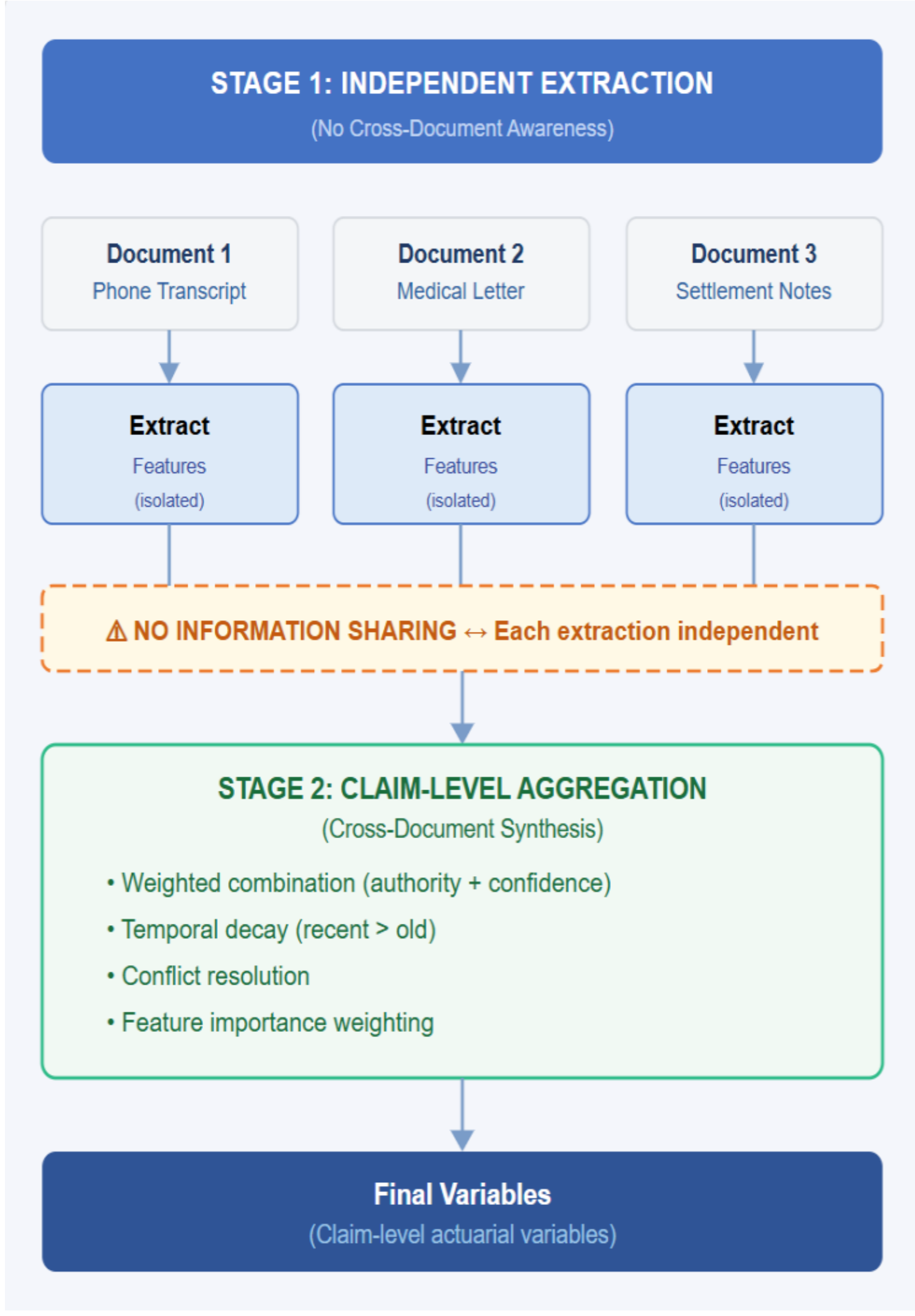

Figure 3. Full prototype workflow.

The preprocessing pipeline includes basic quality validation to filter unsuitable documents. The system enforces a minimum content length threshold (configurable, typically 50+ characters) to avoid processing empty or corrupted files. Documents below the threshold are logged and skipped to prevent wasting LLM API resources. Additional validation during processing checks for proper data structure and sufficient content and checks that the JSON format is well formed and that documents are not empty. This basic quality control ensures processing efficiency while accommodating the variability of real-world document formats.

Stage 1 processes documents sequentially but independently. Each document is analyzed in isolation without awareness of other documents in the claim. This ensures that each document's features are extracted based solely on the content, preventing information bleeding between documents and maintaining clean document-level extractions, as shown in Figure 3.

Stage 2 provides the case context where the temporal nature of claims becomes relevant. Claims naturally evolve over time. Initial injury assessments may differ from later diagnoses, treatment plans change based on patient response, and settlement terms supersede earlier projections. Stage 2 aggregates all Stage 1 extractions together, applying compound scoring that accounts for document weight (authority level), confidence scores, document age and recency (temporal decay), and feature importance (variable-specific weighting):

(1) Document authority weights range from 1.0 for settlement notes (most authoritative final assessments) to 0.4 for phone transcripts (early preliminary information).
(2) Confidence scores from Stage 1 extractions (0–1 scale) directly multiply into the final weight.
(3) Temporal decay applies to older documents when more recent evidence exists.
(4) Variable-specific feature importance weights prioritize relevant document types (e.g., medical provider assessments weighted 0.90 for injury severity vs. 0.38 for phone transcripts).

These four factors combine multiplicatively to produce final feature weights for aggregation. Full technical details on the compound scoring algorithm are provided in TAMD Section 3.3 (“Compound Scoring System”). This approach enables the system to track injury progression across the claim timeline, detect inconsistencies between documents, and systematically resolve conflicts by prioritizing more recent, authoritative, and reliable sources. The two-stage separation ensures clean feature extraction in Stage 1 while enabling sophisticated temporal analysis in Stage 2.

### 2.3.2 Priority integration approach

Documents are assigned priority weights reflecting their authority and reliability for actuarial variable extraction. Settlement documents receive highest priority as final authoritative sources. Medical records and provider letters provide objective clinical assessments. Adjuster reports contribute professional investigation findings. Clinical notes offer ongoing treatment documentation. Claimant statements represent subjective first-person perspectives. These priority weights combine with confidence scores and temporal recency in Stage 2 synthesis to systematically resolve conflicting information across documents.

The document processing strategy integrates seamlessly with the two-stage architecture described in “Document categorization .” Preprocessed documents enter Stage 1 with standardized structure, metadata, and quality validation. Document-specific prompts leverage the identified document type to apply appropriate extraction logic. Quality flags and priority assignments inform Stage 2 synthesis decisions. The result is a robust pipeline that handles real-world documentation variability while maintaining the consistency and reliability required for actuarial analysis.

Detailed preprocessing specifications, including metadata extraction logic, date parsing algorithms, and quality assessment criteria, are documented in TAMD and the code repository.

## 2.4 Use case: Medical records processing

Medical records processing serves as the primary use case demonstrating the prototype's capabilities, addressing the critical gap where valuable clinical information remains untapped due to manual review constraints. The system processes clinical documentation through the two-stage pipeline with specialized extraction prompts. Clinical notes (from FHIR medical encounters) and medical provider letters both undergo Stage 1 extraction targeting clinical severity indicators, treatment complexity assessments, functional status and work restrictions, and recovery trajectory predictions.

Stage 2 synthesis aggregates multiple medical documents within a claim, tracking medical evolution over time. Medical provider letters receive high document weight (0.95–1.00), reflecting authoritative clinical perspective, while clinical notes receive 0.90–0.95 as ongoing assessments.

The medical records use case demonstrates LLM-based processing feasibility and value. By converting narrative clinical information into structured actuarial variables, the prototype enables incorporation of previously inaccessible medical insights into traditional analytical frameworks, with extensibility suggesting broad applicability across claims processing workflows.

## 2.5 Validation and accuracy assessment

The validation methodology establishes how the prototype's extraction accuracy and reliability will be assessed. The validation strategy benchmarks extraction accuracy against synthetic data with known ground truth values. Success is measured through three key metrics: classification accuracy of LLM-derived severity indicators against expert review, consistency of feature extraction across similar claim types measured by interrater reliability scores, and completeness of structured data extraction from unstructured sources measured by successful field population rates in the JSON output schema. This controlled testing environment with known claim characteristics enables rigorous assessment of the framework's performance before application to production data.

### 2.5.1 Validation methodology

The validation strategy employs a multilayered approach assessing different aspects of system performance. Synthetic data validation uses known ground truth from Synthea FHIR-generated test cases to verify that the system correctly extracts expected information when true values are known. This provides baseline confidence in extraction algorithms and identifies systematic errors in processing logic. Expert review benchmarking will compare LLM extractions to human expert assessments on a sample of real claims documents, measuring agreement rates and identifying categories in which human-AI disagreement is most common. Consistency testing evaluates whether the system produces stable results when processing the same documents multiple times, essential for establishing reliability in production environments. Production deployment might require a reproducibility system that is not part of our current prototype.

For this project, we measure success through the classification accuracy of LLM-derived severity indicators against expert review. In the case of unstructured medical notes, experts typically are clinicians whose judgment is valuable but expensive and time-consuming to obtain. Comparing our sample of synthetic data to the results of expert judgment will ensure that automated assessments align with professional standards. We measure consistency of feature extraction across similar claim types with interrater reliability scores, demonstrating the reliability and reproducibility of the approach. We measure completeness of structured data extraction using successful field population rates in the JSON output schema, indicating the system's ability to convert unstructured information into usable actuarial variables systematically.

Three primary metrics assess system performance. Classification accuracy measures the percentage of actuarial variables correctly extracted compared to ground truth or expert assessment. Confidence score calibration evaluates whether the system's self-reported confidence scores (ranging from 0.0 to 1.0) accurately reflect actual extraction accuracy. High-confidence extractions should demonstrate higher accuracy rates than low-confidence extractions. Completeness measures the proportion of required actuarial variable fields successfully populated in output JSON, identifying document types or variable categories where extraction frequently fails or returns insufficient information.

Development-phase validation focused on establishing basic system functionality. The synthetic data generation pipeline (Scripts 1–2) was validated through structural checks confirming that Script 2 accepts only properly formatted Script 1 output, that generated documents maintain internal consistency with underlying medical profiles, and that all required JSON fields are properly populated. Processing pipeline validation confirmed that all four scripts execute successfully with proper error handling, produce schema-compliant JSON output at each stage, and handle the six defined document types without processing failures. Manual spot-checking during development identified obvious errors in prompt engineering and variable extraction logic requiring correction.

### 2.5.2 Quality assurance framework

The validation protocol will measure accuracy by variable category, document type, and claim complexity level. Confidence score calibration will be assessed by comparing self-reported confidence levels against actual extraction accuracy across the validation set. Results inform refinement of prompts, variable definitions, and Stage 2 aggregation logic. The validation procedure proceeds as shown in Figure 4. This procedure was adapted from Tam et al. (2024).

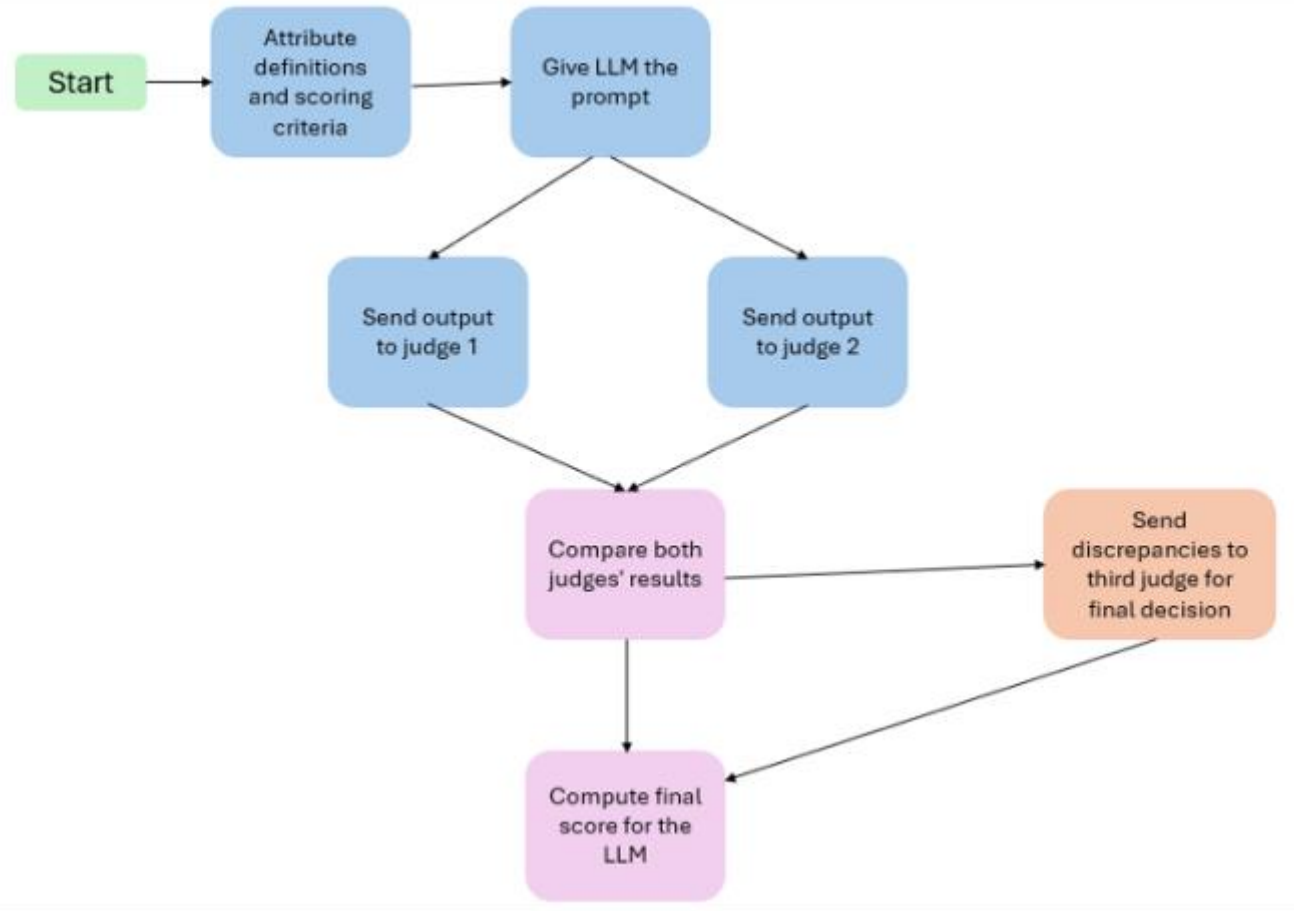

Figure 4. Validation procedure and expert review protocol.

The current study used two independent reviewers to complete the validation exercise shown in Figure 4. The reviewers are clinical experts who are pharmacists with Doctor of Pharmacy (PharmD) degrees. Both are currently in a fellowship training program at a major academic medical center.

The overall approach was a claims-level assessment of the LLM output compared to ground truth as assessed by the reviewers. Documents were used to assess the claims-level accuracy; however, the results of each document produced in Stage 1 were not used in most cases. Instead, reviewers had access to all relevant documents and could choose which ones to use to perform their review.

The reviewers scored the results of the Stage 2 processing of the synthetic claims using a structured rubric and a five-point Likert scale ranking. They ranked 10 quality measures for the LLM output along three dimensions: quality and accuracy, understanding and reasoning, and safety and ethics. The Likert scale ranged from 1 (strongly disagree) to 5 (strongly agree). For example, if the LLM assessed a claim as being “moderate” severity and the graders agreed with the accuracy of that ranking, they would assign a 4 or 5. If the LLM assessed a claim as having a treatment type of “surgical,” but it provided limited justification based on the source documents, the graders would assign a 1 or 2. Claims with low scores or clear hallucinations were assessed to determine whether the prototype could be improved to enhance the output for improvement of the prototype itself.

Finally, the overall values, consistency, and reliability of the raters themselves are assessed using several measures. The mean and standard deviation of the scores are used to assess the overall performance of the model, where higher mean and lower standard deviation within a given rater is indicative of better model performance. Agreement across raters is assessed using weighted kappa and mean absolute difference as interrater reliability measures, where a higher weighted kappa and lower mean absolute difference are indicative of better model performance. The standard deviation likely represents a composition of variability in the reviewer and quality of the system outputs, as discussed further in the “Limitations and Risks” section (Section 5.1).

Score divergence of 2 points or more on the output for a specific claim for a specific measure is indicative of the need to bring in a third rater to resolve the difference. For example, if one grader scored the contextual understanding of the LLM for a given claim as 3 (neither agree nor disagree) while the other grader scored the same claim and dimension as 5 (strongly agree), then one of the project leads could independently arbitrate that disagreement through an audit system that flags the claims and dimensions on which the graders disagree by 2+ points. In these cases, the grades themselves are hidden to allow for an unbiased review by the evaluator.

# 3 FEATURE ENGINEERING IMPLEMENTATION

This section describes the technical implementation of the LLM-based claims analysis prototype. The implementation translates the conceptual architecture into a working system that processes real-world claims documents and extracts actuarial variables.

The prototype consists of a four-script Python pipeline supported by modular infrastructure components. Scripts 1 and 2 generate synthetic test data for validation purposes, while Scripts 3 and 4 form the core processing pipeline for real claims analysis. The system is built using industry-standard technologies and follows software engineering best practices to ensure maintainability and extensibility.

The prototype implements a modular four-script pipeline wherein each script performs a specific function. This design provides flexibility: Organizations can run the complete pipeline for synthetic data testing (scripts 1-4 as shown in Table 2) or use Scripts 3–4 independently for real claims analysis (as seen in the last two rows of Table 2). .

## 3.1 DOCUMENT-SPECIFIC EXTRACTION

Table 2. Four-script processing approach.

| Script | Process | Purpose | Input | Output |
|---|---|---|---|---|
| **1** | FHIR Bundle Processor | Convert synthetic medical data to clinical records | FHIR JSON | JSON with clinical notes |
| **2** | Document Augmentation | Enrich claims with additional synthetic documents | Script 1 JSON | Enhanced JSON with multiple documents |
| **3** | Text Aggregator | Convert raw text to structured format | Raw text files | Structured JSON |
| **4** | Claims Analyzer | Extract actuarial variables (two-stage LLM) | JSON from 2 or 3 | Actuarial variables JSON |

The system supports two distinct operational modes:

**Workflow 1—Synthetic data testing:** FHIR bundles → Script 1 → Script 2 → Script 4 → actuarial variables. This workflow generates synthetic claims data with known characteristics, enabling validation without requiring access to real claims data or patient information.

**Workflow 2—Real claims analysis:** raw text files → Script 3 → Script 4 → actuarial variables. This workflow processes actual claims documents for production actuarial analysis.

## 3.2 PROMPT ENGINEERING APPROACH

**Script 1** processes Synthea-generated FHIR bundles containing synthetic medical data and converts them to clinical notes simulating real medical documentation, enabling testing with known ground truth.

**Script 2** augments claims with additional document types (adjuster notes, phone transcripts, claimant statements, settlement documents) using LLMs to generate realistic synthetic content that matches the claim's medical profile.

**Script 3** converts raw text files (medical records, adjuster notes, correspondence) into structured JSON format, standardizing diverse input formats and identifying document types automatically. This preprocessing enables Script 4 to handle real-world claims data with varying file formats and naming conventions.

**Script 4** implements the core two-stage analysis pipeline. Stage 1 processes each document independently using document-specific prompts to extract preliminary features. Stage 2 aggregates Stage 1 results, resolves conflicts through compound scoring, and generates final claim-level actuarial variables organized by functional category.

**Document-specific extraction**

The system employs document-specific prompts optimized for each document type's format and content. Medical records use prompts focused on clinical terminology, diagnostic reasoning, and treatment complexity. Adjuster reports emphasize liability assessment, investigation findings, and reserve recommendations. This specialization improves extraction accuracy compared to generic approaches that treat all documents identically. However, we have not performed systematic A/B testing with quantitative performance metrics across prompt variations. The template architecture supports controlled experimentation (documented in TAMD Section 10.3, “Prompt Optimization”), and structured prompt evaluation with formal metrics would strengthen production deployment.

**Confidence scoring**

Each extraction includes confidence scoring, documenting the LLM’s certainty about the extracted value. Variables with low confidence scores are flagged for manual review, ensuring quality control while maintaining automated processing efficiency. The system also captures source text snippets supporting each extraction, providing transparency and enabling actuarial verification of automated results. The prototype extracts 14 core variables from the 36-variable taxonomy defined in Table 1. These core variables were selected based on three criteria: immediate predictive value for actuarial analysis, reliable extractability from typical claims documentation, and direct alignment with existing actuarial workflows. The remaining variables in the taxonomy provide a road map for future enhancement.

**Validation aggregation (Stage 2)**

The system uses JSON schemas to define structure, data types, and validation rules for both input and output data. This standardization ensures data quality and facilitates integration with external actuarial systems. The output schema includes claim identifiers as a critical linking mechanism, enabling organizations to augment existing datasets with LLM-extracted insights. This design recognizes that validated numerical values typically exist in transactional systems, while qualitative indicators such as litigation risk and medical complexity remain trapped in narrative text.

## 3.3 Cross-Document Synthesis

The prototype generates JSON files containing extracted actuarial variables, establishing a foundation for integration with existing actuarial workflows. This proof of concept demonstrates feasibility while acknowledging that production deployment requires additional development beyond the prototype scope.

The system writes JSON output files to designated directories following a two-stage structure. Stage 1 outputs contain document-level extractions with confidence scores and evidence. Stage 2 outputs contain claim-level aggregated variables organized by actuarial function (reserving, ratemaking, claims management). Performance speed is about 5–10 minutes per claim. Processing costs are highly model dependent, ranging from $0.50 to $5.00 per claim. Actual costs vary by document complexity.

Organizations implementing the prototype can develop batch processes to periodically parse these JSON files and load variables into their data warehouses or actuarial systems. The standardized JSON format and controlled vocabularies (e.g., “minor,” “moderate,” “major,” or “catastrophic” for injury severity) facilitate programmatic access through Python; R; or database extract, transform, and load tools.

## 3.4 Confidence Scoring and Justification

The prototype implements basic quality controls through configurable confidence thresholds and quality flags. The system applies a default confidence threshold (0.7 for Stage 1, 0.7 for Stage 2) to filter low-quality extractions, generates claim-level quality flags alerting reviewers to processing concerns, and calculates aggregate confidence scores enabling organizations to prioritize claims for manual review.

We use a universal fallback template for unrecognized document types, but extraction quality is lower than for specialized templates. We recommend (1) monitoring fallback template usage frequency, (2) monitoring quality scores by document type, and (3) developing specialized templates when new formats exceed 5% of volume. The modular architecture supports adding templates without code changes (see TAMD 3.2). Organizations should implement validation protocols including confidence score filtering before integration, periodic auditing of LLM extractions against expert samples, and monitoring of variable population rates to identify systematic extraction failures.

# 4 Results and performance

## 4.1 Overall system performance

Claims naturally evolve. Medical records may show initial injury assessments that differ from later treatment outcomes. Adjuster reports capture liability perspectives that may conflict with claimant statements. Settlement documents represent final resolutions that supersede earlier evaluations. The system employs a hybrid thinking approach that achieves chain-of-thought reasoning benefits through structured rationale requirements rather than explicit “think step by step” prompting. Every extraction must include (1) value, (2) detailed rationale explaining determination logic, (3) rationale type (verbatim quote vs. derived/calculated value), (4) source document citation, and (5) confidence score. This structure forces the model to perform explicit reasoning, multistep calculations, source citation, conflict resolution, and confidence calibration, creating a complete audit trail of the decision-making process.

Stage 2 aggregations additionally document cross-document synthesis and conflict resolution (e.g., explaining why settlement amounts override earlier estimates based on authority and recency). By processing documents independently before synthesizing results, the system achieves several key advantages:

**Document specialization** through type-specific extraction logic tailored to medical records, legal documents, and settlement agreements

**Temporal tracking** of claim evolution, with explicit handling of progression and changes over time

**Source reliability weighting** that prioritizes authoritative documents (settlement records, medical records) over subjective sources (claimant statements)

**Complete audit trails,** documenting which documents contributed to each variable, essential for regulatory compliance

**Modular extensibility,** allowing new document types or variables without architectural redesign

This modular design enables the system to handle the heterogeneity of real-world claims data while maintaining consistency and transparency. The architecture balances accuracy with practical usability, adding minimal processing overhead while significantly improving extraction reliability through systematic conflict resolution.

## 4.2Performance metrics

Performance was assessed through independent scoring of the 20 claims used in the prototype by two reviewers as described in section 2.5.2. The overall performance is shown in **Error! Reference source not found.**.

Table 3. Validation scoring summary—independent expert assessment (prior to third-reviewer adjudication).

| Reviewer | Average Score | Standard Deviation |
|---|---|---|
| **1** | 4.01 | 1.14 |
| **2** | 4.22 | 1.25 |

The validation results demonstrate that the LLM extraction system achieves reliable performance across the core variable categories. Both reviewers provided consistent assessments of extraction quality, with average scores greater than 4 on a five-point scale. The relatively low standard deviation (1.14 and 1.25) indicates consistent performance across different claim types and complexity levels. A weighted kappa score for the two reviewers was 0.53, which demonstrates a moderate level of agreement across the raters.

The reviewers had 38 divergent scores out of 200 assessments (38 disagreement items across 17 claims), which represents a 19% divergence rate. This divergence likely reflects expected variability in subjective judgments as well as extraction failures. A total of five claims had three or more disagreements, while the other 12 claims had one or two disagreements across the reviewers. The values in these claims were adjudicated by an independent third reviewer, whose assessment served as the final value for these claims' accuracy.

## 4.3 Demonstration: Enhanced reserving through severity segmentation

We applied LLM-extracted severity classifications to enhance traditional chain ladder reserving methods, demonstrating practical actuarial value from this approach. The illustration here extends the claims developed for the purpose of the prototype development to a synthetic test dataset of 420 claims. These claims were developed with known ground truth ultimate costs ($28,128,598) to showcase how results of an LLM-based record assessment could be used as part of a larger actuarial workflow. We compared two approaches: pooled chain ladder analysis treating all claims identically and severity-segmented chain ladder using LLM-derived claim severity categories.

We performed a severity segmentation in which the results of an analysis are segmented based on categorization of the claims by severity. The segmentation, shown below in Figure 5, reveals the characteristic distribution of workers' compensation claims, with "major" severity claims (92 claims, 21.9% of volume) driving 82.2% of total IBNR. "Moderate" severity claims (158 claims, 37.6% of volume) contribute 16.1% of IBNR, while "minor" claims (170 claims, 40.5% of volume) represent only 1.7% of IBNR. This concentration pattern demonstrates how LLM-extracted severity enables more accurate reserve allocation by recognizing that high-severity claims develop differently than routine cases. This approach is likely to be useful for claim monitoring, resource allocation, and other applications.

Figure 5. Severity-segmented reserve distribution.

SEVERITY BREAKDOWN

### Segmented chainladder results

| SEGMENT | CLAIMS | LATEST PAID | ULTIMATE | IBNR | SHARE OF SEGMENTED IBNR |
|---|---|---|---|---|---|
| Major | 92 | $16,058,199 | $16,746,356 | $688,157 | 82.2% |
| Moderate | 158 | $8,666,700 | $8,801,153 | $134,453 | 16.1% |
| Minor | 170 | $2,598,770 | $2,613,225 | $14,454 | 1.7% |
| Total | 420 | $27,323,670 | $28,160,733 | $837,063 | 100% |

**Minor**
Fast emergence, low noise, minimal tail.

**Moderate**
Steady emergence, balanced noise profile.

**Major**
Longer tail, materially heteroskedastic increments.

The severity-segmented approach also demonstrated improvement in reserve accuracy. By stratifying claims into minor, moderate, and major categories based on LLM extraction, the method produced 2.5% improvement in accuracy for overall IBNR, as shown in Figure 6. The segmented approach yielded an ultimate estimate of $28,160,733 with IBNR of $837,063, compared to the pooled method's $28,180,995 ultimate and $857,325 IBNR. Against the synthetic ground truth of $28,128,598 ultimate and $804,928 IBNR, the segmented method achieved 4.0% error versus the pooled method's 6.5% error. It should be noted we did not measure the improvement here in terms of statistical significance, which would require analysis beyond the scope of this paper.

Figure 6. Chain ladder results by method.

### Method comparison

| METHOD | ULTIMATE | IBNR | IBNR ERROR VS. TRUE |
|---|---|---|---|
| Pooled chainladder | $28,180,995 | $857,325 | 6.5% |
| Severity-segmented chainladder | $28,160,733 | $837,063 | 4.0% |
| Synthetic ground truth | $28,128,598 | $804,928 | 0.0% |

This demonstration establishes proof of concept for integration with existing actuarial methodologies. The severity classifications extracted from unstructured medical records and adjuster notes provide stratification variables that improve traditional loss development techniques without requiring wholesale replacement of established actuarial methods. Further development also could include assessing difference in accuracy by accident year and determining whether incorporating these new features would improve estimates for recent accident years compared to older years.

## 4.4 Error analysis and limitations

Analysis of the validation results identified three primary patterns of extraction uncertainty. First, boundary cases between adjacent severity categories (e.g., moderate vs. major) accounted for approximately 60% of reviewer divergences, reflecting the inherent ambiguity in categorical thresholds rather than systematic extraction failures. Many of the divergences occurred on boundary cases whose categorical distinctions are inherently ambiguous (e.g., distinguishing moderate from major severity in borderline cases).

Second, document types varied in extraction reliability, with formal medical provider letters and settlement documents yielding higher confidence scores (typically 0.85–0.95) compared to informal phone transcripts and initial claimant statements (typically 0.65–0.80). Third, specific variables showed differential performance: Injury severity and medical complexity demonstrated high consistency (>90% reviewer agreement), while prospective indicators like litigation risk and settlement likelihood showed lower confidence scores, appropriately reflecting their inherently predictive nature. However, this performance likely is due to the medical document focus of this study. Litigation risk is a separate issue that would require more domain-specific study in the future.

The confidence scoring system demonstrated appropriate calibration. Variables with confidence scores greater than 0.85 showed >95% agreement with expert assessment, while scores between 0.70 and 0.85 showed approximately 85% agreement, and scores less than 0.70 correctly flagged cases requiring manual review. This calibration enables practical quality control: Organizations can set confidence thresholds based on their risk tolerance and manual review capacity, automatically flagging uncertain extractions while processing high-confidence cases without human intervention.

# 5 Discussion

## 5.1 Limitations and risks

### 5.1.1 Data and domain limitations

This prototype was developed and validated using synthetic claims data generated specifically for testing and validation purposes; ground truth labels and expected outputs were known in advance. While this approach enables systematic evaluation of LLM extraction accuracy and provides controlled testing conditions for framework development, it introduces an important limitation. Performance on real production claims data may differ from results observed with synthetic examples. Synthetic data generation necessarily simplifies the full complexity of actual claims documentation, potentially creating test cases that are more internally consistent, better formatted, and more standardized than the documents actuaries encounter in production environments. LLM-generated text tends to follow predictable patterns that LLMs recognize well. Real medical records might have idiosyncrasies, abbreviations, OCR errors, and ambiguities that synthetic data may not capture.

The LLM framework may perform exceptionally well on synthetic medical records that follow clear clinical documentation patterns but face challenges when processing actual provider notes with idiosyncratic formatting, incomplete information, or documentation styles that deviate significantly from standard practices. This gap between synthetic and production data means that organizations implementing the framework must plan for additional validation and potential refinement when applying it to their actual claims portfolios. In general, another limitation is the inherent uncertainty of issues such as medical condition as well as litigation probability. These types of uncertainty and ambiguity likely cannot be captured in synthetic data and may not be feasibly assessed by an LLM.

Synthetic data may not capture the full spectrum of real-world document quality issues that significantly impact extraction accuracy and reliability. Severe OCR errors from poor-quality scanned documents, in which critical medical terms are corrupted or entire sections are illegible, are difficult to replicate authentically in synthetic generation. Highly nonstandard formatting, such as handwritten margin notes, mixed-language documentation, or unconventional document layouts used by specific providers or legacy systems, may not be represented in synthetic test sets. Extremely dense clinical notes with complex nested information, extensive abbreviations, and specialist-specific terminology present processing challenges that synthetic examples may underestimate.

Time stamps are one other limitation with the system, as we do not explicitly time-stamp to maturity milestones or track initial versus final progression. A production-ready version needs `initial_severity`/`final_severity`, with dates and milestone tracking. While the system currently captures document time-stamps and the settlement schema extracts `settlement_date`, we have not added the cost valuation date as a dedicated field linked to cost values. Given the extraction complexity required, identifying correct evaluation dates from text, handling format variations and missing dates, resolving conflicts when multiple dates appear, and ensuring accuracy meets actuarial data quality standards, we will document this as a high-priority enhancement in TAMD Section 10 ("Extension and Enhancement") (see Appendix C).

For production deployment, organizations must conduct validation against actual claims data from their specific portfolios, implementing appropriate deidentification and privacy protections to comply with HIPAA and other regulatory requirements. This real-world validation should assess performance across the full range of document quality, formatting variability, and content complexity encountered in production, with particular attention to edge cases and challenging document types wherein synthetic data may have provided overly optimistic performance estimates.

The current prototype operates under the assumption that input data have been pre-deidentified prior to LLM processing, allowing the research to focus on core extraction methodologies rather than implementing comprehensive privacy protection systems. This design choice means the prototype does not include HIPAA-compliant processing safeguards, comprehensive audit logging for protected health information access, or data residency controls required for production deployment in regulated insurance environments. While this approach is appropriate for research and proof-of-concept development, organizations seeking to implement the framework in production must recognize that significant additional infrastructure is required to meet legal and regulatory obligations. The prototype provides the analytical foundation for LLM-based claims processing but deliberately excludes the privacy and security architecture that production systems require, acknowledging that implementation details vary substantially across organizations based on their specific regulatory context, risk tolerance, and existing technology infrastructure.

### 5.1.2 Prototype limitations

The current prototype demonstrates integration feasibility but does not implement production-grade features such as comprehensive error handling for missing or low-confidence extractions, automated retry logic for failed API calls, or real-time monitoring dashboards. The results of the chain ladder method are also illustrative, and the differences shown are not statistically significant and would need to be tested with the use of actual data in the chain ladder package or other software for further actuarial development (Casualty Actuarial Society 2025).

Organizations implementing the prototype should start with file-based integration and limited claims volume to validate extraction quality before scaling. The modular architecture supports this incremental approach, enabling organizations to build confidence in the technology while managing implementation risk.

## 5.2 Next steps and implementation

### 5.2.1 Integration infrastructure requirements

Organizations adopting the framework must develop additional infrastructure including automated extract, transform, and load workflows for systematic data loading, real-time monitoring and alerting systems to detect processing failures or quality degradation, comprehensive error handling and retry logic to manage transient failures and edge cases, and audit trails for regulatory compliance documenting data access and processing decisions. The prototype’s file-based output provides flexibility for organizations to implement integration approaches suited to their technical architecture and operational requirements, whether through batch processing, database staging tables, or API-based consumption.

### 5.2.2 Privacy and production readiness

Production deployment necessitates robust deidentification methodologies that remove or mask protected health information before documents enter LLM processing workflows, secure cloud infrastructure with appropriate access controls and encryption for data at rest and in transit, and comprehensive privacy impact assessments that evaluate risks specific to the organization's claims portfolio and operational environment. Organizations must implement audit trails documenting which data were processed, when, and by whom, ensuring compliance with regulatory requirements and enabling investigation of potential privacy incidents. Beyond initial deployment, ongoing model validation represents a critical but often underestimated requirement for production systems. Claims documentation language evolves as medical terminology changes, new treatment modalities emerge, and organizational documentation practices shift. Regulatory requirements for AI systems continue to develop, with potential new obligations for explainability, fairness testing, and bias detection. The LLM framework requires periodic revalidation to ensure that extraction accuracy remains acceptable as the linguistic patterns and document characteristics in production data drift from the original training and validation sets. Organizations must budget for continuous monitoring of model performance, periodic validation studies, and framework updates to maintain reliable operations over time.

## 5.3 Future research

This report details a prototype method for processing unstructured claims data using LLMs. We see several additional forms of future research that can extend both the capabilities of this system and its validity. Those include additional types of data and lines of insurance, adoption within real-world systems for insurance, and extension of the validation method for a wider range of insurance and risk management tasks that can be automated with an LLM.

While medical records processing serves as the primary proof of concept, the framework's two-stage architecture naturally extends to other document types and insurance lines beyond the medical-driven claims demonstrated here. Future research should systematically evaluate the framework's performance across property damage assessments in homeowners and commercial property insurance, where photographic evidence and contractors' estimates supplement narrative documentation. Opportunities include extracting litigation risk indicators, third-party involvement patterns, documentation describing business operations, and incident response measures that could inform both underwriting and claims handling. Each expansion requires careful consideration of domain-specific terminology, development of appropriate variable taxonomies, and validation against subject matter expert judgment in that line of business.

Additional types of media and data beyond the text-based claims documentation used here also are an area of future research. Audio and video evidence will be a key area used to enhance claim assessment beyond key metrics that can be captured by humans. This also would extend to image files (e.g., X-rays and photographs) and handwritten notes that may be part of medical records. Tones in conversations; human behaviors indicating fraud, waste, and abuse; and damage that is hidden either due to the limits of human judgment or due to intentional behavior are important areas for future research.

The transition from prototype demonstration to production deployment in real insurance operations presents significant research opportunities spanning technical integration, organizational workflow design, and change management. Future work should address the practical challenges of integrating LLM-extracted variables into existing actuarial modeling platforms, API specifications, and validation protocols. Research is needed on human-AI collaboration models that define appropriate roles for automated processing versus expert review, including decision frameworks for determining when confidence scores or complexity indicators should trigger human oversight. The economic case for LLM adoption requires rigorous measurement of processing cost reductions, improved reserve accuracy, and faster claims cycle times compared to manual approaches. Finally, the actuarial profession may need validated protocols for explaining LLM-derived insights to regulators, senior management, and external auditors, addressing concerns about model transparency, bias detection, and the appropriate level of human oversight in automated decision systems.

Another opportunity is research on cross-review among LLMs to see the variance between different LLMs as well as variations among different types of prompts in the same LLM. Both are active areas of current research, as the training and performance of existing LLM models vary widely both in general and on specific tasks. Reprompting or adjusting prompts within an LLM model has been shown to produce different outputs and different levels of output quality (Haase et al. 2025). Applying these evolving areas of research to the insurance domain is an important area of future research.

The validation framework developed here establishes a foundation for evaluating LLM performance on actuarial tasks. Using synthetic data with known ground truth followed by expert comparison of real data is an appropriate test case, but broader applications of automation in insurance and risk management require more sophisticated validation approaches. Future research should develop standardized benchmarks for evaluating LLM performance across the full spectrum of actuarial tasks, from straightforward data extraction to complex reasoning about reserve adequacy or pricing appropriateness, with particular attention to edge cases and adversarial scenarios that test system robustness.

Calibration of confidence scores, characterization of systematic error patterns, and appropriate propagation of extraction uncertainty through downstream actuarial calculations are three types of improvements that would enhance the prototype model. The use of retrieval augmented generation (RAG) and other techniques to improve the prompting of LLM models and enhance LLM outputs may improve accuracy beyond the version shown here, and that would be an ideal area for future research as RAG technologies become more standardized and integrated into LLM tools. In a secure data environment, additional documents and data could include insurance policy details and the insurer’s policy and claim histories.

Finally, validation methodologies should extend beyond technical performance metrics to encompass usability from a human-centered design perspective, practitioner trust and interpretability, and an appropriate override of automated recommendations when professional judgment indicates concerns with system outputs. The extension and application of these techniques to the prototype work described here will ensure impact through the safe and reliable application of LLM technologies to solve key insurance and actuarial business problems.

# 6 Appendices

## A. Technical specifications and architecture (detailed)

The architecture of the LLM-based claims analysis system is defined by a rigorous multilayered schema architecture coordinating data validation and processing across a two-stage analysis pipeline. This design ensures consistency, quality control, and the preservation of clear data lineage throughout the process.

The validation structure encompasses three distinct schema layers:

**Layer 1: Input validation schemas,** which verify document conformance, encoding formats, and minimum content requirements before processing begins, effectively rejecting malformed inputs early

**Layer 2: Stage 1 document-level schemas,** which define the structured output format for extractions from individual documents

**Layer 3: Stage 2 claim-level schema,** which defines the final aggregated output structure that serves as the integration contract with external actuarial systems

The core processing operates across two stages, beginning with Stage 1, document-level extraction. This stage is focused on analyzing each document independently to extract document-specific features, maximizing fidelity to the source text. The extraction is managed through a sophisticated system utilizing category templates for high-level guidance, specialized templates tailored for specific document types (like clinical notes or adjuster notes), function schemas that define the target JSON output structure, and function calling to enforce compliance with these schemas via the LLM API. The output of this stage for each document includes the extracted variable values, a calculated confidence score (0.0–1.0) indicating extraction certainty, and a detailed rationale explaining how the value was derived, thereby establishing a per-document audit record.

Following Stage 1, Stage 2 performs cross-document aggregation. The purpose of this stage is synthesis: combining all the individual document extractions for a single claim to produce a unified set of final actuarial variables. This synthesis must systematically resolve conflicting information, which is achieved through the compound scoring system. This system ranks sources based on a four-component weighted formula: document weight (authority of the source, e.g., medical notes rank higher than initial phone transcripts), confidence score (LLM extraction certainty), temporal recency (reflecting that recent information is generally more reliable), and feature importance (variable-specific document relevance).

The final outputs are structured JSON files defined by the Stage 2 claim-level schema, which organizes the variables into three functional categories—reserving, ratemaking, and claims management—directly aligning with typical actuarial workflows. To ensure transparency and auditability, the final output includes comprehensive provenance tracking, linking the final variable assessments back to the specific Stage 1 source documents and providing explicit explanations of any conflict resolution decisions. Quality Indicators are generated at the claim level, serving to flag cases that require manual actuarial review due to factors such as low confidence or unresolved source conflicts, thus shifting quality control from preventive input filtering to detective output validation. This structured JSON output enables automatic validation and provides the essential integration contract for consumption by external actuarial systems.

# B. Code repository guide

The codebase follows a clear directory structure:

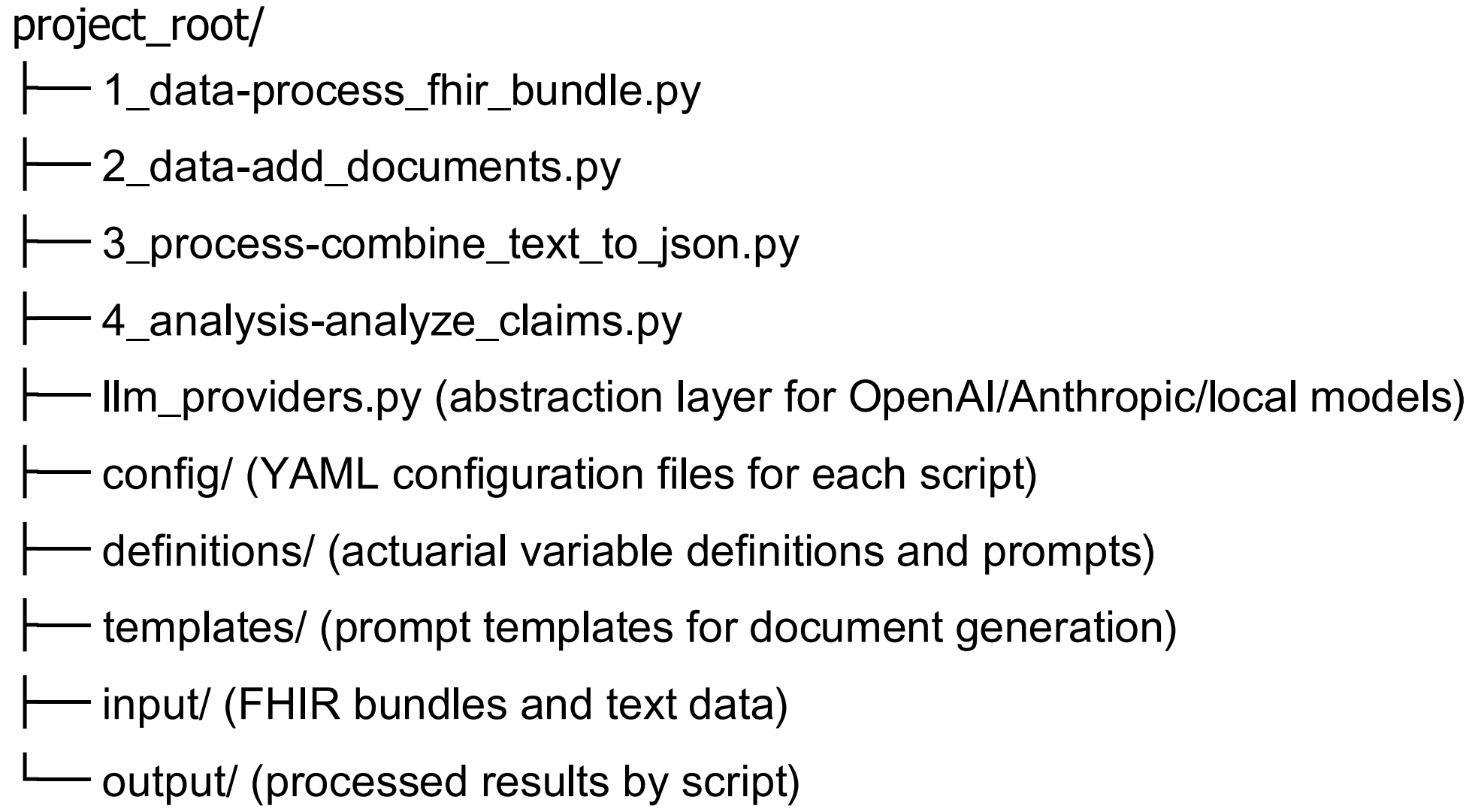

```
project_root/
├── 1_data-process_fhir_bundle.py
├── 2_data-add_documents.py
├── 3_process-combine_text_to_json.py
├── 4_analysis-analyze_claims.py
├── llm_providers.py (abstraction layer for OpenAI/Anthropic/local models)
├── config/ (YAML configuration files for each script)
├── definitions/ (actuarial variable definitions and prompts)
├── templates/ (prompt templates for document generation)
├── input/ (FHIR bundles and text data)
└── output/ (processed results by script)
```

This modular organization enables independent development and testing of each component while maintaining clear data flow and version control. All configuration is externalized through YAML data serialization files, allowing operational adjustments without code changes.

The system is built on Python with careful technology selection balancing capability, cost, and accessibility for actuarial organizations.

**Core technologies:**

- **Python** 3.12+ as the primary programming language, providing extensive data science libraries and broad organizational familiarity
- **OpenAI GPT-4o mini** as the primary LLM provider for production-quality extraction accuracy
- **Ollama** integration enabling local model deployment for organizations with data residency requirements

**Key Python libraries:**

- **OpenAI** for LLM API interaction
- **PyYAML** for configuration management
- **Pydantic** for JSON schema validation

The system supports multiple deployment configurations. Organizations can use cloud-based LLM APIs (OpenAI) for rapid deployment with minimal infrastructure or deploy local models via Ollama for complete data control and elimination of per-token costs. The provider abstraction layer in llm_providers.py enables switching between providers without modifying analysis code.

For cloud deployment, minimal infrastructure is needed: Python environment, API credentials, and network access to LLM providers. Processing scales with API rate limits rather than local compute capacity. For local deployment, GPU acceleration is recommended for acceptable performance with open-source models (Llama 3, Qwen 2.5), though CPU-only operation is possible for small-scale testing. Cost considerations for the prototype focused on research and validation. Production deployments at scale should evaluate local model options to control costs, particularly for high-volume processing.

## C. Technical Architecture and Methodology Document (TAMD)

This appendix is available online at https://www.casact.org/sites/default/files/2026-03/Appendix-C-Technical-Architecture.pdf.

## D. Technical specification requirement

# CAS LLM—Technical Specification Requirement

*Appendix D to: Leveraging LLMs for Unstructured Claims Data Analysis*

*The Casualty Actuarial Society Artificial Intelligence Working Group provided funding and support for this research.*

## D.1. CAS LLM—Technical Specification Requirement

- **Date:** October 1, 2025
- **Status:** Requirements Specification (Prototype)
- **Project:** Leveraging LLMs for Unstructured Claims Data Analysis
- **Target Audience:** CAS stakeholders and researchers
- **Scope:** Research prototype for LLM claims analysis
- **Architecture:** Modular four-script pipeline
- **Code Repository:** `https://github.com/mdsight/llm-claims-analysis`

## D.2. Executive Summary

This specification defines the technical requirements for a prototype system that processes unstructured claims text using OpenAI's LLM API to extract actuarial variables. The system will consist of four Python scripts: synthetic data generation (Scripts 1 and 2) for testing and core claims analysis (Scripts 3 and 4) for processing. These scripts will follow a local Python processing workflow with optional cloud storage integration, designed to demonstrate feasibility and provide a foundation for future research and implementation.

The prototype will process claims text through a secure pipeline with robust error handling and comprehensive logging for audit trails to ensure reliable extraction of actuarial variables. Key actuarial variables produced by the prototype will include variables for reserving, ratemaking, and claims management. Security measures will include the use of pre-anonymized synthetic data and service account authentication for cloud-based services where applicable.

## D.3. System Requirements

### D.3.1. Pipeline Architecture

- **Synthetic Data Generation (Testing)**
  - Script 1: FHIR medical bundles → JSON with clinical notes
  - Script 2: Script 1 output → Enhanced JSON with synthetic documents
- **Core Claims Analysis (Main Function)**
  - Script 3: Raw text files → Structured JSON preprocessing
  - Script 4: JSON input → Two-stage LLM analysis → Final actuarial variables
- **Script Dependencies**

– Scripts 1, 2, and 3 can run independently
– Script 4 requires input from Script 3 *or* Scripts 1 and 2
– Minimum working system: Scripts 3 and 4

### D.3.2. Core Functionality Requirements

- **Script 3—Text Preprocessing**
  - Batch-process patient folders with text files and metadata
  - Support six document types: `phone_transcript`, `medical_provider_letter`, `adjuster_notes_initial`, `settlement_adjuster_notes`, `claimant_statement`, `clinical_note`
  - YAML metadata: `name`, `age`, `gender`, `patient_initials`
  - Batch process multiple patient folders with basic error handling
- **Script 4—LLM Analysis**
  - Stage 1: Extract document-specific features with confidence scores
  - Stage 2: Aggregate information across documents with priority weighting
  - Output standardized actuarial variables in JSON format
- **Scripts 1 and 2—Synthetic Data Generation**
  - Script 1: Convert FHIR bundles to JSON with clinical notes
  - Script 2: Enhance with synthetic phone transcripts and adjuster notes
  - Create test cases with known outcomes for validation

### D.3.3. Input Requirements

- **Script 1 Input—FHIR Bundles**
  - **Format:** JSON files containing Synthea FHIR medical bundles
  - Must include medical encounters and clinical notes
  - Standard FHIR R4 format compliance
- **Script 2 Input**
  - JSON output files from Script 1
  - Must contain clinical notes and patient information
- **Script 3 Input**
  - **Folder structure**

```
input/text_data/
├── patient_AB/
│   ├── AB_metadata.txt
│   ├── AB_ClaimID1_phone_transcript.txt
│   ├── AB_ClaimID1_medical_provider_letter.txt
│   └── AB_ClaimID1_adjuster_notes_initial.txt
└── patient_CD/
    ├── CD_metadata.txt
    ├── CD_ClaimID2_phone_transcript.txt
    └── CD_ClaimID2_settlement_adjuster_notes.txt
```

  - **Metadata file format (YAML in .txt file)**

```
name: Mary Rodriguez
age: 45
gender: female
patient_initials: mr
```

- **Script 4 Input**
  - JSON output from Script 1, 2, or 3

- Structured format with patient data and document content
- **Input Validation Requirements**
  - UTF-8 text encoding for all documents
  - Maximum individual document size of 10 MB
  - Required metadata fields must be present
  - Document files must have content (not empty)

### D.3.4. Output Requirements

- **Core Actuarial Variables**
  - **Reserving variables**
    - `Injury_Severity_Category`: [`Minor/Moderate/Major/Catastrophic`]
    - `Medical_Complexity`: [`Simple/Moderate/Complex/Highly_Complex`]
    - `Treatment_Type`: [`Conservative/Surgical/Ongoing/Complex_Rehab`]
    - `Ultimate_Cost_Category`: [`<25K/25K-75K/75K-150K/>150K`]
  - **Ratemaking variables**
    - `Risk_Level`: [`Low/Medium/High/Extreme`]
    - `Causation_Type`: [`Equipment/Environmental/Human_Factor/Multiple`]
    - `Industry_Risk_Category`: [`Construction_Heavy/Construction_Light/Other`]
    - `Safety_Compliance`: [`Full/Partial/None/Unknown`]
    - `Experience_Modifier_Impact`: [`Favorable/Neutral/Adverse`]
  - **Claim management variables**
    - `Litigation_Risk`: [`Low/Medium/High`]
    - `Settlement_Likelihood`: [`High/Medium/Low`]
    - `Management_Complexity`: [`Routine/Moderate/Complex/High_Touch`]
  - **JSON Output Format**

```
{
  "claim_id": "string",
  "processing_timestamp": "ISO 8601 timestamp",
  "reserving_variables": {
    "Injury_Severity_Category": {
      "value": "Major",
      "confidence_score": 0.85,
      "source_documents": ["medical_provider_letter"]
    }
  },
  "claim_management_variables": { "...similar structure..." },
  "confidence_scores": {
    "overall": 0.82
  },
  "processing_metadata": {
    "encounters_processed": 3,
    "input_source": "text_derived"
  }
}
```

## D.4. Technical Implementation

### D.4.1. Technology Stack

- **Runtime:** Python 3.12+
- **LLM Service:** OpenAI API (GPT-4o mini)
- **Configuration:** YAML files for settings
- **Storage:** Local file system only, with optional cloud storage integration
- **Output:** JSON files with actuarial variables

### D.4.2. Environment and Dependencies

- **System Requirements**
  - **Operating system:** Windows, macOS, or Linux
  - **Hardware:** Standard laptop (8GB RAM recommended)
  - **Internet:** Required for OpenAI API access
- **Dependencies**
  - **Python:** Python 3.12+
    - **Core**
      - `openai`—OpenAI API client for LLM integration
      - `pyyaml`—YAML configuration file processing
      - `argparse`—Command-line argument parsing (built-in)
      - `pathlib`—Path handling (built-in)
      - `json`—JSON processing (built-in)
      - `re`—Regular expressions (built-in)
      - `logging`—Logging functionality (built-in)
      - `sys`—System functionality (built-in)
      - `os`—Operating system interface (built-in)
      - `datetime`—Date and time handling (built-in)
      - `time`—Time-related functions (built-in)
      - `copy`—Object copying utilities (built-in)
      - `base64`—Base64 encoding (built-in)
    - **Cloud Integration (optional)**
      - `google-cloud-storage`—Google Cloud Storage integration
  - **Authentication:** Environment variables for API keys

### D.4.3. Configuration and Deployment

- **Configuration Management**
  - YAML files for processing parameters
  - Document type settings and priorities
  - API rate limiting settings
  - Quality control thresholds
- **Deployment Approach**
  - Local execution only, with optional cloud integration
  - Standard Python script execution
  - File-based input/output
  - Command-line interface
- **Processing Modes**
  - **Standard:** Full pipeline (Script 3 → Script 4)

- **Development:** Individual script execution for testing
    - **Validation:** Synthetic data workflow (Scripts 1 → 2 → 4)
- **Planned Optional Features**
    - **Cloud:** Google Cloud Storage integration
    - **LLM provider:** LLM abstraction layer for extensibility
        - Support for alternative models and providers
    - **UI framework:** Modular architecture enabling future UI development

### D.4.4. Security and Quality Control

- **Basic Security**
    - Store API keys in environment variables
    - Process only synthetic/anonymized data
    - No persistent storage of sensitive information
- **Quality Control**
    - Confidence scoring for all extractions
    - Error handling with continued processing
    - Basic input validation
    - Comprehensive logging for troubleshooting

## D.5. Validation and Extensions

### D.5.1. Testing Requirements

- **Functional Testing**
    - All scripts execute without errors
    - Script outputs compatible between stages
    - Synthetic data produce expected results
    - Manual testing with sample claims
- **Performance Expectations**
    - Process individual claims in a reasonable amount of time (under 5 minutes each)
    - Handle small batches
    - Acceptable for research/prototype use

### D.5.2. Research Validation

- **Basic Validation Methods**
    - Compare outputs with known synthetic data results
    - Manual review of sample extractions
    - Consistency testing with repeated processing
- **Validation Framework**
    - Framework supports testing different prompts/approaches
    - Easy modification of variable definitions
    - Extensible to additional document types

### D.5.3. Research Extensions

- **Future Research Capabilities**
  - Additional document types beyond the initial six
  - New actuarial variables and definitions
  - Different LLM models and approaches
  - Extension to other insurance lines
- **System Extensibility**
  - Modular design supports adding new components
  - Configuration-driven approach enables easy modifications
  - Clear interfaces between processing stages

### D.5.4. Success Criteria

- **Functional Success**
  - All four scripts successfully execute
  - Produces JSON output with required actuarial variables
  - Processes both synthetic and real text files
  - Basic error handling prevents crashes
- **Research Success**
  - Demonstrates LLM capability for claims analysis
  - Provides framework for further research
  - Generates usable synthetic test data
  - Enables comparison of different approaches
- **Prototype Acceptance**
  - Working demonstration of core concept
  - Documentation enables other researchers to use system
  - Extensible design for future enhancements
  - Reasonable processing time for research purposes